\documentclass[aps,pre,twocolumn,amssymb,amsmath,floatfix]{revtex4}
\usepackage{graphicx,subfigure}
\usepackage{bm}
\usepackage{verbatim}
\usepackage{amsmath}
\usepackage{amssymb}
\usepackage[T1]{fontenc}
\usepackage{ae,aecompl}

\newcommand{\rv}{{\bf r}}
\newcommand{\av}{{\bf a}}
\newcommand{\cv}{{\bf c}}

\newcommand{\qv}{{\bf q}}

\newcommand{\nh}{{\hat n}}

\newcommand{\xh}{{\hat x}}
\newcommand{\yh}{{\hat y}}
\newcommand{\zh}{{\hat z}}
\newcommand{\eh}{{\hat e}}

\newcommand{\oh}{{\frac{1}{2}}}

\newcommand{\cH}{{\mathcal H}}
\newcommand{\bse}{\begin{subequations}}
\newcommand{\ese}{\end{subequations}}
\def\rf#1{(\ref{#1})}
\def\rfs#1{Eq.~\rf{#1}}

\begin{document}

\title{Nonlinear smectic elasticity of helical state in cholesterics 
and helimagnets} 
\author{Leo Radzihovsky}
\affiliation{Department of Physics, University of Colorado, Boulder, CO 80309}
\author{T. C. Lubensky}
\affiliation{Department of Physics, University of Pennsylvania,
  Philadelphia, Pennsylvania 19174}
\date{\today}

\begin{abstract}
  General symmetry arguments, dating back to de Gennes dictate that at
  scales longer than the pitch, the low-energy elasticity of a chiral
  nematic liquid crystal (cholesteric) and of a Dzyaloshinskii-Morya
  (DM) spiral state in a helimagnet with negligible crystal symmetry
  fields (e.g., MnSi, FeGe) is identical to that of a smectic liquid
  crystal, thereby inheriting its rich phenomenology. Starting with a
  chiral Frank free-energy (exchange and DM interactions of a
  helimagnet) we present a transparent derivation of the fully
  nonlinear Goldstone mode elasticity, which involves an analog of the
  Anderson-Higgs mechanism that locks the spiral orthonormal
  (director/magnetic moment) frame to the cholesteric (helical)
  layers. This shows explicitly the reduction of three orientational
  modes of a cholesteric down to a single phonon Goldstone mode that
  emerges on scales longer than the pitch. At a harmonic level our
  result reduces to that derived many years ago by Lubensky and
  collaborators\cite{cholestericLubensky}.
\end{abstract}

\maketitle

Dating back to the original cholesterol liquid crystal discovered by
Reinitzer, chirality plays a central role in modern study of liquid
crystals\cite{DP}. It is equally important for understanding helical
states of noncentrosymmetric magnets (e.g.,
MnSi)\cite{Ishikawa,BakJensen}, driven by a chiral
Dzyaloshinskii-Morya (DM) interaction\cite{DM}

Among a wealth of induced phenomena\cite{DP} chirality converts 
uniform nematic and ferromagnetic phases into states in which the
orientational (nematic or spin) field twists periodically, thereby
leading to a variety of spatially modulated phases, such as the
cholesteric, two- and three-dimensional ``Blue''
phases\cite{2dBlue,WrightMermin,DP} and recently discovered Skyrmion
line crystals in MnSi helimagnet\cite{skyrmionMnSi}. These
spontaneously break the translational and rotational symmetries,
forming liquid-crystalline structures that are periodic along $1$, $2$
or $3$ dimensions. A cholesteric and helical phases are the most
ubiquitous of these, characterized by a biaxial order with
\begin{eqnarray}
  \nh_0(\rv)
  &=&\xh_1\cos(\qv\cdot\rv)+\xh_2\sin(\qv\cdot\rv),
\end{eqnarray}
breaking the translational symmetry along a single, spontaneously
selected axis, $\xh_3\equiv\zh$, where $\xh_1,\xh_2,
\xh_3\equiv\xh_1\times\xh_2$ form an orthonormal triad that is
constant in the ground state. 

General symmetry arguments\cite{DP} and an explicit derivation at a
{\em harmonic} level\cite{cholestericLubensky} applied to this
spontaneously layered helical state predict that at scales longer than
the helical pitch, the low-energy (Goldstone mode) elasticity is
identical to that of a smectic liquid crystal\cite{DP}. The three
orientational degrees of freedom defining the cholesteric at short
scales thereby reduce to a single smectic-like phonon mode, that
emerges on scales longer than the pitch.  The associated enhanced
fluctuations in\cite{GP,GW} and disordering of these states leads to
rich phenomenology\cite{DP,CL}, and in the case of MnSi is believed to
be associated with the striking observation of the non-Fermi liquid
behavior\cite{nonFL}.

Although by now quite familiar, the symmetry breaking of the helical
state falls outside the conventional $G/H$ paradigm\cite{CL}. Despite
fully breaking the $O(3)$ rotational symmetry of the Euclidean group
the state is characterized by only a {\em single} $U(1)$ Goldstone
mode, $\chi$, the spiral's phase related to the smectic phonon
$u=-\chi/q_0$. As we will show below, the absence of the two
additional orientational Goldstone modes is best understood as a
mathematical equivalent of the Anderson-Higgs
mechanism\cite{DP,DunnLubensky,CL} that gaps them out.

In this Letter we explicitly show how this single low-energy helical
mode emerges and derive its nonlinear smectic energy functional,
expected from the underlying rotational symmetry\cite{DP,CL}. Because
the latter leads to harmonic phonon fluctuations that diverge in three
dimensions and below, the inclusion of nonlinearities is essential for
a sensible and self-consistent description, as anticipated long ago by
Lubensky, et al.,\cite{cholestericLubensky} and Grinstein and
Pelcovits\cite{GP,GW}. With the neglect of crystalline anisotropy all
of our cholesteric results apply equally well to the description of
the low-energy bosonic modes of the helical state in the DM
helimagnets such as MnSi\cite{Ishikawa,BakJensen}.

The helical texture $\nh_0(\rv)$ minimizes the chiral Frank-Oseen
free-energy density of a chiral nematic\cite{DP,CL,FrankHcomment}
\bse
\begin{eqnarray}
\cH&=&\oh K_s(\nabla\cdot\nh)^2 + 
\oh K_b(\nh\times\nabla\times\nh)^2,\nonumber\\
&&+ \oh K_t(\nh\cdot\nabla\times\nh+q_0)^2,\\
&=&\oh K\left[(\partial_i\nh_j)^2+2q_0\nh\cdot\nabla\times\nh +
  q_0^2\right],
\label{H*_F}
\end{eqnarray}
\ese
where in the second form we focused on the isotropic limit,
$K_s=K_b=K_t\equiv K$ and dropped the total derivative saddle-splay
(Gaussian curvature) contribution, that reduce the Hamiltonian to that
of DM ferromagnet in the absence of crystal symmetry breaking
fields\cite{DM,BakJensen}.  Within this approximation the space-spin
($\rv-\nh$) coupling only enters through the chiral (second) term,
with the elastic (first) piece explicitly exhibiting independent
rotational invariances of space, $\rv$ and of the director, $\nh$
(spin in the MnSi magnet context).


We now look at long-wavelength, low-energy Goldstone modes excitations
about the helical ground state, $\nh_0(\rv)$. A general state is
described by
\begin{eqnarray}
  \nh(\rv)&=&\eh_1(\rv)\cos(\qv\cdot\rv+\chi(\rv))
  +\eh_2(\rv)\sin(\qv\cdot\rv+\chi(\rv)),\nonumber\\
\label{nGeneral}
\end{eqnarray}
where $\eh_1(\rv),\eh_2(\rv),
\eh_3(\rv)\equiv\eh_1(\rv)\times\eh_2(\rv)$ now constitute a spatially
dependent orthonormal frame, describing the orientation of the local
director (spin) helical plane, that is independent of the helical axis
set by $\qv\cdot\rv+\chi(\rv)=const.$ The phase $\chi(\rv)=-q_0
u(\rv)$ of the chiral helix
also defines the phonon field $u(\rv)$ of the helical layers, which on
scales longer than the pitch $a = 2\pi/q_0$ define the smectic-like
displacement of these helical phase fronts. Thus, altogether on the
intermediate scales there are three independent orientational degrees
of freedom, $\chi(\rv)$ and $\eh_3(\rv)$. The azimuthal angle
$\phi(\rv)$, defining the orientation of the $\eh_{1,2}$ around
$\eh_3$ is redundant to $\chi(\rv)$, as it can be eliminated in favor
of it via a local gauge-like transformation on $\eh_{1,2}$. Although
naively, the low-energy coset space is isomorphic to $(S_1\times
S_2)/Z_2=RP_3$,
as we will see below, only a single Goldstone mode, $\chi(\rv)$ will
survive this helical symmetry breaking.

Substituting $\nh(\rv)$ from \rfs{nGeneral} into the free-energy of
the chiral nematic (helimagnet), \rfs{H*_F}, and using
\begin{eqnarray}
\partial_i\nh_j&=&(q_i+\partial_i\chi)[-\eh_{1j}\sin(\qv\cdot\rv+\chi)
+\eh_{2j}\cos(\qv\cdot\rv+\chi)]\nonumber\\
&&+\partial_i\eh_{1j}\cos(\qv\cdot\rv+\chi)
+\partial_i\eh_{2j}\sin(\qv\cdot\rv+\chi),
\label{epsn_r}
\end{eqnarray}
together with 
\bse
\begin{eqnarray}
\partial_i\eh_{1j}&=&a_i\eh_{2j} + c_{1i}\eh_{3j},\\
\partial_i\eh_{2j}&=&-a_i\eh_{2j} + c_{2i}\eh_{3j},
\label{identies}
\end{eqnarray}
\ese
and gauge field ``spin-connections''
\bse
\begin{eqnarray}
a_i&=&\eh_{2}\cdot\partial_i\eh_{1},\\
c_{1i}&=&-\eh_{1}\cdot\partial_i\eh_{3},\\
c_{2i}&=&-\eh_{2}\cdot\partial_i\eh_{3},
\label{connections}
\end{eqnarray}
\ese
we find
\begin{widetext}
\bse
\begin{eqnarray}
\cH&=&\frac{K}{2}\bigg[(a_i\eh_{2j} + c_i\eh_{3j})\cos(\qv\cdot\rv+\chi)
+ (-a_i\eh_{1j} + c_{2i}\eh_{3j})\sin(\qv\cdot\rv+\chi)\nonumber\\
&& + (q_i+\partial_i\chi)\big[-\eh_{1j}\sin(\qv\cdot\rv+\chi)
+ \eh_{2j}\cos(\qv\cdot\rv+\chi)\big]\nonumber\\
&& +q_0(\eh_{2i}\eh_{3j}-\eh_{3i}\eh_{2j})\cos(\qv\cdot\rv+\chi)
+q_0(\eh_{3i}\eh_{1j}-\eh_{1i}\eh_{3j})\sin(\qv\cdot\rv+\chi)\bigg]^2
,\hspace{.5cm}\\
&=&\frac{K}{2} (\nabla\chi + \av + \qv - q_0\eh_{3})^2
+ \frac{K}{4} (\cv_1 + q_0\eh_{2})^2
+ \frac{K}{4} (\cv_2 - q_0\eh_{1})^2
\label{Hchi}
\end{eqnarray}
\ese
\end{widetext}
In obtaining Eq.\rf{Hchi}, we dropped the constant and oscillatory
parts, that average away upon spatial integration.

We first note that the requirement of well-defined helical phase
fronts, i.e., the absence of dislocations and disclinations in the
layer structure, can be enforced by the compatibility condition
$\nabla\times\av=0$, consistent with Mermin-Ho
relation\cite{MerminHo}. This allows us to take $\av=\nabla\phi$
and thereby shift (``gauge'') away $\phi(\rv)$ in favor of
$\chi(\rv)$, according to $\chi(\rv)+\phi(\rv)\rightarrow\chi(\rv)$.

Without loss of generality, we next take $\qv=q_0\zh$
with $\zh$ defining the orientation of the helical axis in the
laboratory coordinate system $\xh,\yh,\zh$. The long-wavelength
free-energy density, reexpressed in terms of the smectic-like phonon
field $u(\rv)$ and the local nematic helical frame orientation $\eh_3$
then reduces to
\begin{eqnarray}
\cH&=&\frac{K q_0^2}{2}(\nabla u +\eh_{3}-\zh)^2
+ \frac{K}{4} (\cv_1^2 + \cv_2^2)\nonumber\\
&&+ \frac{K q_0}{2} (\cv_1\cdot\eh_{2} - \cv_2\cdot\eh_{1}).
\label{Hchi2}
\end{eqnarray}
Clearly the first term, above, accounts for the energetic cost of the
deviation of the local nematic frame $\eh_{3}$ from the local
orientation of the helical layers. A minimization of this term (or
equivalently at long wavelengths, in a statistical mechanical
treatment integrating out the independent $\eh_{3}(\rv)$ degree of
freedom), at low-energies locks the orientations of the cholesteric
layers and the nematic frame. In a perturbative treatment this leads
to the expected relation
\begin{equation}
\nabla_\perp u \approx - \eh_{3\perp},
\end{equation}
that is an example of a Higg's-like mechanism (akin to thermotropic
smectic liquid crystals\cite{deGennes,DP,CL}), that at long scales
effectively gaps out the orientational Goldstone modes.

The exact minimization over the unit vector $\eh_3(\rv)$ can also be
carried out using a Lagrange multiplier $\lambda$ to impose the
constraint $\eh_3\cdot\eh_3=1$. Minimization over $\eh_3(\rv)$ gives
\begin{eqnarray}
\nabla u +\eh_{3}-\zh = -\lambda\eh_3,
\label{minimizeLambda}
\end{eqnarray}
with the solution 
\bse
\begin{eqnarray}
\lambda + 1 &=& \sqrt{(\nabla u -\zh)^2}
=\sqrt{1-2u_{zz}},\label{lambda}\\
\eh_3&=&(\zh-\nabla u)/\sqrt{1-2u_{zz}},\label{eh3}
\end{eqnarray}
\label{lambdaeh3}
\ese 
where 
\begin{eqnarray}
u_{zz}=\partial_z u - \oh(\nabla u)^2
\label{uzz}
\end{eqnarray}
is the standard fully nonlinear smectic strain tensor, that encodes
the full rotational invariance of the helical state\cite{GP,CL}.
Using Eqs.\rf{lambdaeh3} to eliminate $\lambda$ and $\eh_3(\rv)$ in
favor $u_{zz}$ (valid at long scales), we find
\begin{widetext}
\bse
\begin{eqnarray}
  \hspace{-0.5cm}\cH&=&
  \frac{K q_0^2}{2}(\sqrt{1-2u_{zz}}-1)^2+\frac{K}{4}
  \left[(\eh_{1}\cdot\partial_i\eh_3)(\eh_{1}\cdot\partial_i\eh_3)
    +(\eh_{2}\cdot\partial_i\eh_3)(\eh_{2}\cdot\partial_i\eh_3)\right]
  + \frac{K q_0}{2}
  \eh_{1i}\eh_{2j}(\partial_i\partial_j u - \partial_j\partial_i u),
\label{result1}\\
  &\approx&\frac{B}{2} u_{zz}^2
  + \frac{\overline{K}}{2}(\partial_i\partial_j^\perp u)^2,\label{result2}\\
&\approx&\frac{B}{2}\big[\partial_z u - \frac{1}{2}(\nabla u)^2\big]^2
  + \frac{\overline{K}}{2}(\nabla_\perp^2 u)^2,
\label{Hchi3}
\end{eqnarray}
\ese
\end{widetext}
where to obtain our main result, \rfs{Hchi3} we used the condition of
well-defined helical layers with no dislocations in $u(\rv)$, i.e., a
single-valued phase field $\chi(\rv)$, neglected the boundary terms,
expanded to lowest order in the nonlinear strain tensor $u_{zz}$, and
defined the compressional and bending elastic moduli
\begin{eqnarray}
B&=& K q_0^2,\ \ \ \overline{K}= K/2.
\end{eqnarray}

Thus, as advertised, we have demonstrated that on scales longer than
the helical pitch $2\pi/q_0$, the low-energy deformations (Goldstone
modes) of the helical state are characterized by a fully rotationally
invariant, nonlinear smectic elastic theory\cite{DP,GP,CL}. The latter
can be derived by spontaneously ordering the density $\rho(\rv)$ of
the isotropic fluid into a one-dimensional periodically modulated
state (smectic), characterized by $\rho(\rv)=\rho_0 +
\rho_{q}\cos[(\qv\cdot\rv- q_0 u)]$\cite{GP}. Alternatively, the above
nonlinear compressional form (first term in \rf{result1}) emerges
directly from the de Gennes' gauge theory of the smectic\cite{DP},
after condensing $\rho_q=|\rho_q|e^{i\Phi(\rv)}$ to give
\begin{eqnarray}
\cH_{sm} = \oh B(\nabla\Phi - \nh)^2 + \oh K(\nabla\cdot\nh)^2,
\end{eqnarray}
and then minimizing over the unit director $\nh(\rv)$ as in
\rf{minimizeLambda}-\rf{uzz}.

We note that as required, $\cH$ in Eq.\rf{result1} is a function of
the fully nonlinear strain $u_{zz}$ as it must to preserve full
rotational invariance. Furthermore, it is a nonlinear function of this
strain, that reduces to the familiar ``harmonic nonlinear''
form\cite{GP} in Eq.\rf{Hchi3} only for small $u_{zz}$.  It is worth
observing that through the introduction of the phase field $\Phi = -z
+ u$, this nonlinear in $u_{zz}$ term in \rf{result1} can be written
as $(|\nabla \Phi| - 1)^2$.  This form has been used in recent
analyses of various nonlinear properties of smectic.\cite{KamienSm}

Our result in \rfs{Hchi3} then in turn implies that the helical state
inherits all the novel nonlinear elastic effects previously discovered
in the context of conventional, thermotropics and lyotropic smectic
liquid crystals and other spontaneously layered states that emerge
from an isotropic state. These include thermal
fluctuations\cite{smecticSq,GP,GW} and heterogeneous\cite{RTprb}
anomalous elasticity effects, the undulation
instability\cite{ClarkMeyer}, and many others.

One important distinction from a conventional smectic, however, is the
underlying chirality of the helical layered state.  Although as in the
chiral smectic the effective Anderson-Higgs mechanism expels the
expression of chirality (e.g., twist) inside the helical state (as the
magnetic flux density [twist of the vector potential] is expelled from
the Meissner state)\cite{deGennes,DP,CL}, inclusion of the chiral
terms (encoded in the departure from the $\nabla\times\av=0$ condition
used to get to \rf{Hchi2}) is essential to understanding the
topological defects and melting of the helical
state\cite{RTunpublished}.

We acknowledge the financial support through the NSF DMR-1001240 (LR)
and NSF-DMR-0804900 (TL).


\begin{thebibliography}{99}
\bibitem{FrankHcomment} This is a valid description on intermediate
  and longer scales, where the magnitude of the nematic order
  parameter $Q$ can be taken to be a constant.
\end{thebibliography}
\end{document}